\documentclass{aastex631}
\usepackage{makecell}
\usepackage{booktabs}
\usepackage{graphicx}
\usepackage{float}
\usepackage{comment}
\usepackage{ulem}
\usepackage{color}
\usepackage{url}

\newcommand{\ech}{\'{e}chelle\,}

\begin{document}

\title{Multiple regular frequency spacings in $\delta$ Scuti stars from automated analysis of TESS oscillation spectra}

\author[0000-0003-2896-1471]{Shravan M. Hanasoge}
\affiliation{Tata Institute of Fundamental Research, Mumbai 400005, India}
\correspondingauthor{Shravan Hanasoge}
\email{hanasoge@tifr.res.in}
\author[0009-0007-0535-3074]{Shashwat Chamoli}
\affiliation{Tata Institute of Fundamental Research, Mumbai 400005, India}
\author[0009-0009-1287-6521]{Subrata Kumar Panda}
\affiliation{Indian Institute of Astrophysics, Bengaluru 560034, India}
\affiliation{Tata Institute of Fundamental Research, Mumbai 400005, India}
\author[0000-0003-4854-7550]{Daniel R. Reese}
\affiliation{LIRA, Observatoire de Paris, Universit\'{e} PSL, Sorbonne Universit'{e}, Universit'{e} Paris Cit'{e}, CY Cergy Paris Universit'{e}, CNRS, 92195 Meudon, France}

\begin{abstract}
The oscillation spectra of rapidly rotating $\delta$ Scuti stars are difficult to interpret because rotation modifies both the frequencies and spatial structure of the modes. Although theoretical studies predict that acoustic modes should organize into approximately equally spaced frequency sequences, systematic searches for such patterns have been limited by the absence of scalable, statistically robust detection methods. We present an automated statistical framework for identifying approximately equally spaced mode sequences in stellar oscillation spectra while quantifying the significance of each candidate detection. We apply this method to oscillation spectra of 4,681 $\delta$ Scuti stars observed by the \textit{Transiting Exoplanet Survey Satellite} (TESS). The analysis identifies regular mode sequences in a large number of stars and, in many cases, reveals multiple independent regular frequency spacings within the same oscillation spectrum. While the existence of regular sequences is broadly consistent with theoretical expectations for rapidly rotating acoustic modes, the occurrence of multiple dissimilar spacings is not readily explained by current models. Differences in acoustic travel times between known island-mode families appear insufficient to account for the observed diversity of spacings, suggesting that additional mode families or other physical effects contribute to the organization of oscillation spectra of $\delta$ Scuti stars. Our results provide a scalable framework for studying regular mode sequences in large stellar surveys and new observational constraints on oscillations in rapidly rotating stars.
\end{abstract}


\section{Introduction} \label{sec:intro}

Solar-like stars are generally slow rotators and are nearly spherical in shape. Additionally, they have radiative interiors and cores and thick outer convection zones \citep{Kippenhahn_1990, seismo_resourceful_2010}. Turbulence is most vigorous in the near-surface layers, just below the photosphere where the density drops significantly and the gas becomes optically thin. Turbulent convective flows at the surface become transsonic and undergo shocking, stochastically driving a spectrum of acoustic $p$ modes \citep{Goldreich1977ApJ}. The amplitudes of the oscillations are assigned according to a uniform random process; a balance between the excitation power profile, wave damping and mode inertia results in an overall Gaussian-like envelope for the oscillation spectrum \citep{KjeldsenBedding, Chaplin2005, Kjeldsen2011}. Acoustic modes in these systems are well described by the asymptotic theory of stellar oscillations \citep{Tassoul}, and the mode frequencies are distributed according to a specific and unique frequency spacing. Further complications ensue in terminal-age main-sequence stars where the core g-modes couple with p-modes, resulting in pulsation forests; however, these too are straightforward to interpret and are very well explained by the asymptotic theory of stellar oscillations \citep{Mosser2012}.

 With increasing stellar mass, many of these ``nice" properties start to go away. The interior of a typical intermediate-mass main-sequence star ($1.5\le M/M_\odot \le 2.5$) comprises a convective core, radiative envelope and a thin convective layer at the surface \citep{seismo_resourceful_2010}. A primary consequence is that mode excitation is achieved through the opacity-driven $\kappa$ mechanism (\citealt{kp_mechanism}) {and an unknown non-linear saturation process 
 which results} in unpredictable pulsation amplitudes. This implies that regular sequences of {mode frequencies,}
 a crucial detection representing the start of asteroseismic analysis, may not always be detected. The large separation in non-rotating or slowly rotating stars is the difference in frequencies of modes of consecutive radial order but with the same $\ell$ and $m$. {In the high-frequency range, it is close to $1/(2\int_0^R dr/c_s)$ which is known to scale as the square root of the star mean density.}
 
 An additional complication is the rapid rotation that this class of stars exhibits. Spin rates can approach Keplerian break-up speeds, at which point centrifugal forces overcome self-gravity, resulting in stellar material being flung out \citep{Maeder2009}. Away from this extreme regime, rapid rotation leads to substantial shape distortion (oblateness), with the equatorial layers bulging out and the poles collapsed inward \citep{vanBelle2012}. When considering models of acoustic oscillation in rapidly rotating stars, the classical definition of the large separation can be extended. It has been shown that, most, if not all, high-frequency acoustic modes belong to a series of modes whose spatial structure only differs by their number {of nodes between the surface and the center (\citealt{Lignires2006, Reese2006, Lignires2008, Reese_2008, Lignieres2009, Reese2009, Pasek2011, Reese2017, evano19}).} 
 The large separation in this scenario is again defined as the difference between the frequencies of modes of such series at consecutive {node number.}
{Acoustic modes of deformed stars can be classified from the study of acoustic rays. This approach has been validated for high-frequency modes \citep{Lignieres2009, Reese2009, Pasek2011, evano19} and still provides a relevant framework in lowest frequency ranges of the acoustic modes of $\delta$ Scuti stars \citep{Lignires2006, Reese2006, Reese2017}.
Accordingly, acoustic modes of deformed stars belong to three different families, namely the whispering gallery modes, the island modes and the chaotic modes. The whispering-gallery modes are similar to the modes of a non-rotating star in that their oscillation cavity lies between the surface and a concentric caustic. But they are unlikely to be detected in rapidly rotating stars because they have a large number of regularly spaced nodes on the star surface, strongly limiting their amplitudes in the disc-averaging light curve.
Island modes concentrate around particular acoustic rays that close on themselves, called stable periodic orbits. A stable periodic orbit with only two reflections at the surface exists for all deformations. The associated modes are called 2-period island modes and their number increases with rotation. 
Their large frequency separation is given by the
inverse of the sound travel time along the closed path of the ray $1/\oint \left(ds/c_s\right)=\Delta/2$. The stability of periodic orbits with larger sound travel time and thus the existence of the associated island modes strongly depend on the rotation rate. For example, a stable periodic orbit with 6 surface reflections and associated 6-period island modes are present near  $\Omega= 0.6 \Omega_K$ \citep{Lignieres2009}. Their large separation has the same expression but for a path with 6 reflections. In stars, the sound travel time is strongly dominated by the time spent in the 
near-surface layers, which are the coolest, thereby possessing the lowest sound speeds.  The large separation thus essentially depends on the number of surface reflections of the periodic orbit so that it is very close to $\Delta/6$ for the 6-period island modes.
Chaotic modes are made of chaotic ray paths. These modes also belong to series regularly spaced in frequency, providing a {\it pseudo-large separation} which is close to $\Delta$ \citep{evano19}.
These authors also found a $\Delta/3$ frequency spacing between chaotic modes although this feature has only been found in a small rotation range near $\Omega=0.7 \Omega_K$. It should be noted that this particular regular spacing was not anticipated from the ray dynamics because 
it is due to partial barriers of the chaotic dynamics which are not simple to detect.
To summarize, theoretical studies have shown that the most visible high-frequency modes in deformed stars belong to
series of regularly spaced frequencies with spacings very close to fractions of $\Delta$, the
series with $\Delta$ or $\Delta/2$ being by far the most frequent.
\citet{Reese2017} analyzed acoustic spectra in lowest frequency ranges typical of $\delta$ Scuti stars and also found the $\Delta$ and $\Delta/2$
regular spacings. It must be reminded though that the exploration of theoretical acoustic spectra throughout the HR domain covered by $\delta$ Scuti stars is not complete yet. For example, low frequency spectra of evolved deformed intermediate-mass stars have not been computed.}

{Regular frequency spacings interpreted as large frequency separations have been previously detected in the oscillation spectra of some $\delta$ Scuti stars \citep{GarcaHernndez2009, Bedding_2020, singh2025seismicconstraintsspinevolution}. In particular, \citet{GarcaHernndez2009} considered $\delta$ Scuti stars in binary systems which made it possible to verify that as expected the detected spacings scaled with the square roots of the star mean density.} 
Here, we study oscillations in 4,681 $\delta$ Scuti stars in the TESS catalog \citep[identified by][]{singh2025seismicconstraintsspinevolution}. We find that a dominant fraction of the sample shows regular oscillations over the spacing range $4\le\Delta f\le14$ cy$\cdot{\rm d}^{-1}$. The ratios of the highest-to-lowest amplitude modes in these regular sequences can vary substantially, from $O(1) - O(10^2)$. This large dynamic range makes it challenging to examine each spectrum individually, and we therefore automate this search and identification process using an algorithm described subsequently.

In this work, we present an automated statistical framework for identifying approximately equally spaced mode sequences in stellar oscillation spectra while quantifying the significance of each candidate detection. We apply the method to the oscillation spectra of 4,681 $\delta$ Scuti stars observed by TESS, enabling a homogeneous survey of regular mode sequences on this scale. In addition to identifying numerous candidate stars exhibiting regular frequency sequences, the analysis reveals that many spectra contain multiple independent regular frequency spacings. While the presence of regular sequences is broadly consistent with theoretical expectations for acoustic modes in rapidly rotating stars, the coexistence of multiple dissimilar spacings is not readily explained by current models. The methodology introduced here provides a scalable approach for analyzing large stellar oscillation datasets, while the observational results furnish new constraints on the organization of oscillation modes in rapidly rotating $\delta$ Scuti stars.

\section{Data}
Among the stars observed during the first 63 sectors of TESS, \cite{singh2025seismicconstraintsspinevolution} identified 5381 $\delta$ Scuti pulsators by analyzing their short-cadence (2 minute) oscillations. They observed uniformly spaced modes in the spectra of only 300 $\delta$ Scuti stars, at a rate similar to that recovered by \cite{Bedding_2020} (60 $\delta$ Scuti stars among a sample of $\sim$1,000). \cite{singh2025seismicconstraintsspinevolution} reported the characteristic $\Delta\nu$ values for the 300 stars. The detection of regular frequency patterns is easier in $\delta$ Scuti stars that are in the near Zero Age Main Sequence (ZAMS) phase of evolution \citep{GarcaHernndez2009}. {This may be due to the fact that \it{g} modes increasingly populate the acoustic spectra as the star evolves, leading to avoided crossings with \it{p} modes and thus perturbing their regular spacing.}

Both \citet{Bedding_2020} and \citet{singh2025seismicconstraintsspinevolution} focused on regular frequency patterns comprising high-amplitude modes. In this work, we attempt to discover regular frequency patterns comprising modes with fainter amplitudes, challenging to discern at first glance. {We restricted the sample to stars with masses between 1.5 and 2.5 M$_\odot$ and effective temperatures between 6,300 K and 8,500 K, corresponding to the typical parameter ranges of $\delta$ Scuti stars. This resulted in a final sample of 4,681 $\delta$ Scuti stars, which we subjected to further analysis.}

The procedure to identify regular sequences, outlined by, e.g., \cite{Bedding_2020}, requires manual visualization of the \'echelle diagrams constructed with $\Delta f$ continuously varying in a specified range, only flags high-amplitude modes, does not quantify the false positive rate (which is possible even in that amplitude range), and requires tedious and careful examination of each star. Additionally, since $\delta$ Scuti stars show multiple regular spacings, each spectrum must be studied thoroughly to exhaustively identify all possible sequences. Consequently, we devised an automated approach to identify all sequences present in a spectrum; this method has one major user-prescribed parameter, the threshold false-positive rate (FPR) of detection (which will be described in Section~3) and two minor parameters, the acceptable global and local SNR thresholds for peak identification. {The FPR discussed above refers to the likelihood of emergence of equally spaced frequencies among a set of pulsation modes, purely due to chance.}

\section{Algorithmic approach to detecting sequences} \label{algorithm}
The goal here is to identify a sequence of modes whose frequencies can be described by an approximately linear function. To build a robust technique to achieve this, we need multiple components: (A) extract meaningful oscillation peaks from spectra, (B) determine the parameters with which to represent a mode sequence, (C) construct an algorithm to detect sequences and (D) develop a means to quantify the false-positive detection rate (FPR). The first step is to set a threshold signal-to-noise ratio that defines the cutoff between signal and noise; we conservatively choose sufficiently low values so as to allow in as many peaks as possible and subsequently apply additional filters to retain only consequential peaks (A). We then use the \texttt{find\_peaks} routine from the \texttt{SciPy} signal-processing package to extract the mode frequencies which we then filter for meaningful modes (frequencies are given by $\nu_i$, with $i\le N$, $N$ being the number of selected modes). We also retain amplitude information of all these peaks (B). 

We then parameterize mode sequences as approximate arithmetic progressions (AAPs), i.e., where the frequencies are given by $\nu_{j+1} = \nu_j + \Delta f + \varepsilon_j$, where $\{\nu_j\}$ belong to a specific chain of length $k\ll N$, $\Delta f$ the associated spacing, and $\varepsilon_i$ the local deviation from an exact arithmetic progression (C). We set $|\varepsilon_i| \le \tau$, where $\tau$ is a tolerance level. To detect AAPs at specified tolerances and for given chain lengths, we develop a graph-theory based algorithm (D). The algorithm takes the following flow:
\begin{enumerate}
\item Prescribe a global threshold signal-to-noise ratio (${\rm SNR_{global}}$) and extract frequencies of all modes whose amplitudes are larger than this value. In the present analysis, the noise level is estimated as the median amplitude of the full spectrum, and the SNR of each peak is calculated as the ratio of the peak amplitude to this noise level. Peaks whose SNRs exceed the prescribed global threshold are retained.
\item From the set of peaks retained by the global SNR selection, we further restrict the sample to the 200 peaks with the largest amplitudes. For each such candidate peak at frequency $f$, we define a window of width $\Delta f_{\rm local}$ centered on $f$, excluding a smaller interval of width $\Delta f_{\rm excl}$ around the peak itself. The local noise level is then estimated as the median amplitude of the remaining peaks within this window. The local SNR of the candidate peak is computed as the ratio of its amplitude to this local noise estimate, and only peaks  with ${\rm SNR_{local}}$ exceeding a user-prescribed threshold are retained. Let the number of identified modes after this step be $N$.
\item Modes are sorted in increasing order by their frequencies and we compute their consecutive differences, producing an array of length $N-1$. We denote each element of this difference vector by $\delta_i$. 
\item The sequence-identification problem is reformulated using graph theory. Each point in the difference vector is a node and the path connecting pairs of nodes is an edge. Specifically, we are interested in ``directed graphs"; the direction in this case extends from a node at lower frequency to a counterpart at higher frequency.
\item We determine the paths between all pairs of nodes in this sequence whose edge lengths are equivalent to a number in the range $(\Delta f -\tau)\le\Sigma_i\delta_i \le (\Delta f +\tau$). This provides a list of all mode pairs which are separated by a specific regular spacing to within the allowed tolerance. $\Delta\nu$ is sampled densely in the range $4-14$ $d^{-1}$.
\item The method identifies directed graphs of a minimum user-prescribed length among these pairs. For instance, suppose the modes $(A,B), (B, D), (D, F), (G, H), (I,J)$ have spacings that satisfy the pairwise constraint (5). We see that the directed graph $A\rightarrow B\rightarrow D\rightarrow F$ forms a chain of 4 modes whose frequencies are in AAP. However, $G\rightarrow H$ and $I\rightarrow J$ are two independent pairs that satisfy constraint (4) but are not part of a connected chain.
\item After all candidate chains are identified, chains with $\Delta f$ values that are very close (within $5\%$ of each other) and whose frequencies are subsets of longer chains are removed. Throughout this analysis, we adopt the following parameter values:
\begin{itemize} 
  \item ${\rm SNR_{global}} = 2.0$,
  \item $\Delta f_{\rm local} = 2.0 \, d^{-1}$, 
  \item $\Delta f_{\rm excl} = 0.05 \, d^{-1}$, 
  \item ${\rm SNR_{local}} = 1.2$.
\end{itemize}
\end{enumerate}
The parameter ${\rm SNR_{global}} = 2$ is set this low \citep[as compared to the limits set by][]{charpinet10,zong16}, so as to allow in as many peaks into the full frequency series. We rank order these and only retain the 150 highest-amplitude modes in the analysis. We adopt this conservative approach so as to liberally identify sequences - some of which may be discarded upon further inspection. In this manner, we are unlikely to reject ``good" sequences where one or two modes are of low amplitudes. Lastly, this approach paves the way for selecting sequences where one or two modes within the sequence may be missing or possess undetectably low amplitudes.
Modes are allowed to belong to multiple sequences and we do not enforce uniqueness considerations here.

The final component we need is a means of quantifying a false positive rate (FPR) associated with an AAP - this plays a crucial role in building confidence in the identifications. Intuitively, we expect that the likelihood of finding a sequence of modes in AAP increases with $N$, tolerance $\tau$, and with decreasing chain length $k$. In other words, for a very large number of modes, it is likely that some of the modes will be ordered according to an AAP. Secondly, when the tolerance $\tau$ increases, more chains fall into the acceptable category and the FPR rises. Finally, it is much more likely that short-chain AAPs are present in a series than long-chain AAPs. We have not examined in detail whether the selected modes could be combination frequencies, originating from non-linear interactions between pulsation modes. However, we do not expect them to be so, because the combination frequencies will likely not form regular frequency patterns, especially those comprising 5-6 modes.

Mode amplitude is also an important ingredient in this analysis because high-amplitude short-chain sequences are likely to be more meaningful than low-amplitude sequences of similar length. Ideally, the confidence we place in an identification must codify this aspect. However, we would need a model for the oscillation amplitudes in order to take this into account accurately, which we do not possess. Additionally, $\delta$ Scuti stellar oscillation spectra have widely differing features, some dense with modes and others sparse; busy spectra have a large number of peaks rising above the threshold amplitude, thereby increasing the false positive rate. In comparison, the false-positive rate associated with the detection of AAPs in sparse spectra will be smaller. Thus, some amplitude information is already incorporated into the algorithm, and while the false-positive rate determination can be tuned to scrutinize the amplitudes even more carefully, such efforts require assumptions about the excitation process, which in turn reduce the accuracy of this measure. A second feature that we might consider penalizing for, or alternatively, rewarding, is the consistency of amplitudes in a sequence. The expectation is that modes of a specific sequence are associated with a resonant cavity, are excited by the same process, and are therefore imbued with similar amplitudes. In addition, visual inspection also favors sequences with modes of comparable amplitudes. Although a uniformity criterion can also be taken into account when computing the false positive rate, there is limited understanding of the excitation processes of island modes, and we choose not to include this effect.

We obtain estimates of false-positive rates by performing Monte Carlo simulations. Specifically, we construct a grid over the parameters $(\tau, N, k)$, where $N$ is the number of detected modes and $k$ is the minimum chain length. In the present analysis, $N$ is varied between $40$ and $150$ in steps of 1, $k$ is taken in the range $4-8$ in steps of 1, and $\tau$ is sampled between $0.04$ and $0.11$ $d^{-1}$ in steps of $0.01$. For each point on this grid, we generate a random sequence of ``frequencies" whose values are drawn uniformly between $5$ and $100$ $d^{-1}$, and apply the same sequence-detection algorithm as used for the observations. We search in the range $4 \leq \Delta f \leq 14$ $d^{-1}$ for regular spacings; for each choice of parameters, we run 1,000 Monte Carlo simulations.

\begin{figure*}[!h]
    \centering
    \includegraphics[width=0.4\linewidth]{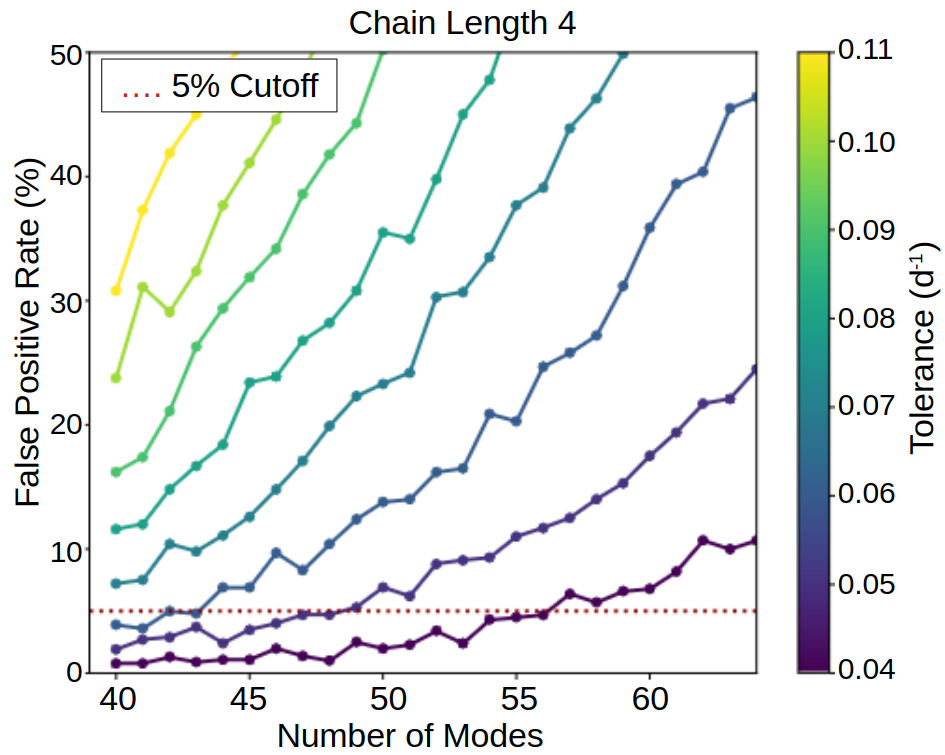}    
    \includegraphics[width=0.4\linewidth]{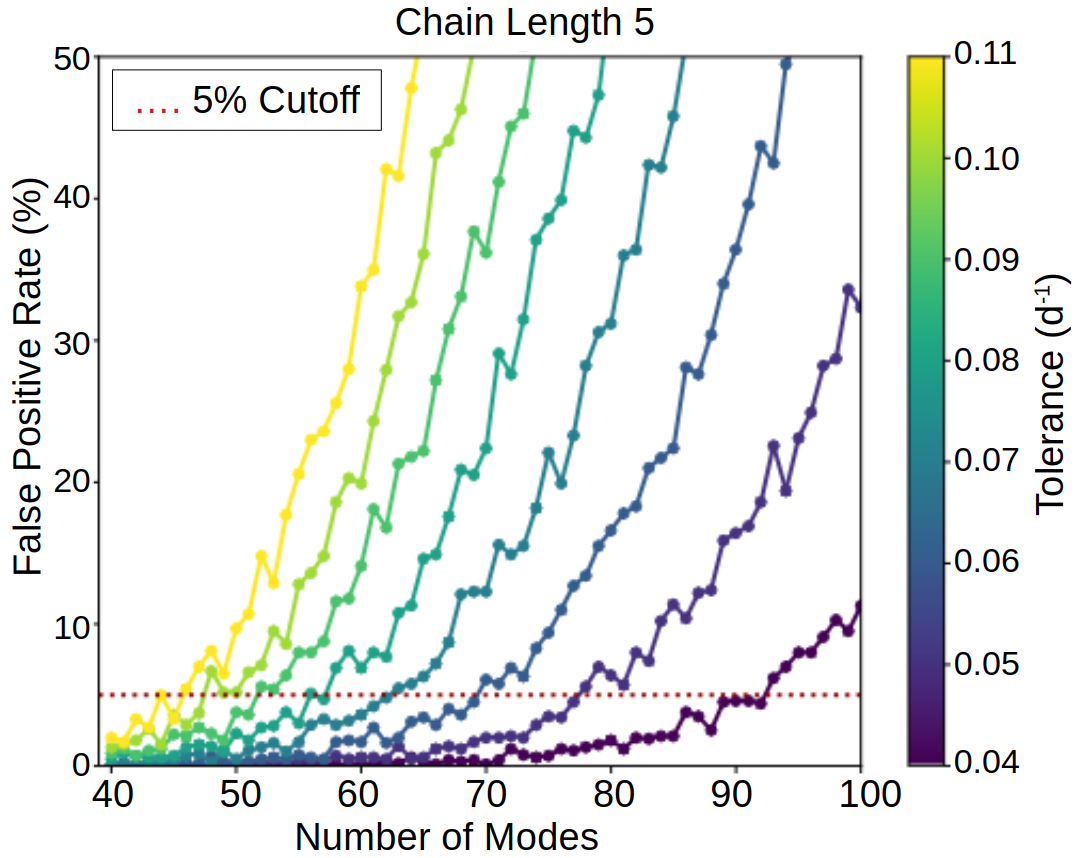}
    \includegraphics[width=0.4\linewidth]{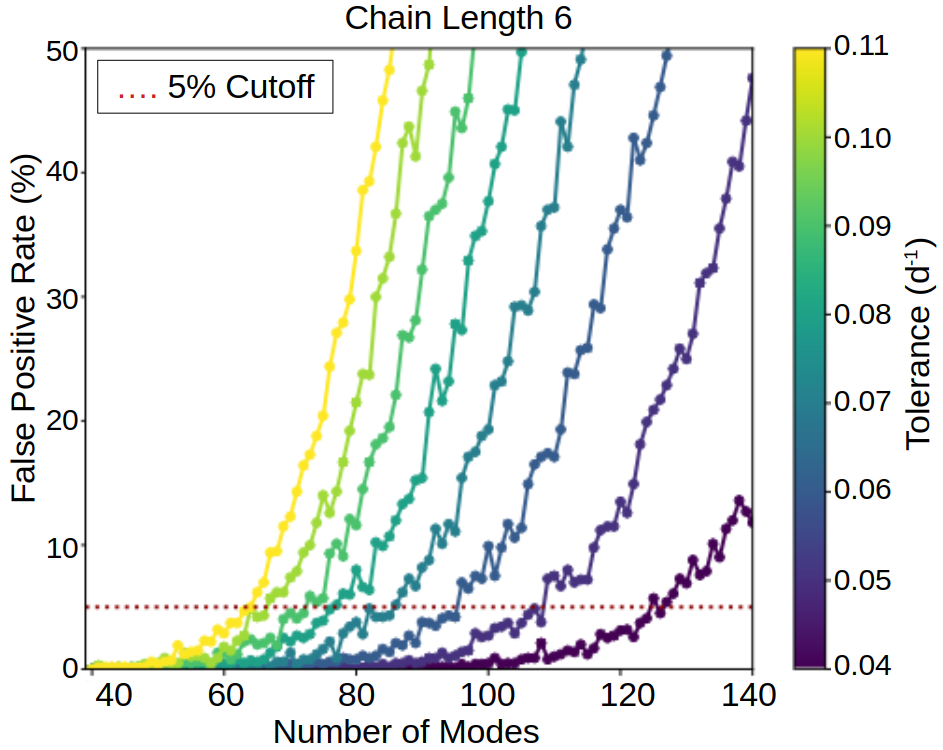}
       \includegraphics[width=0.4\linewidth]{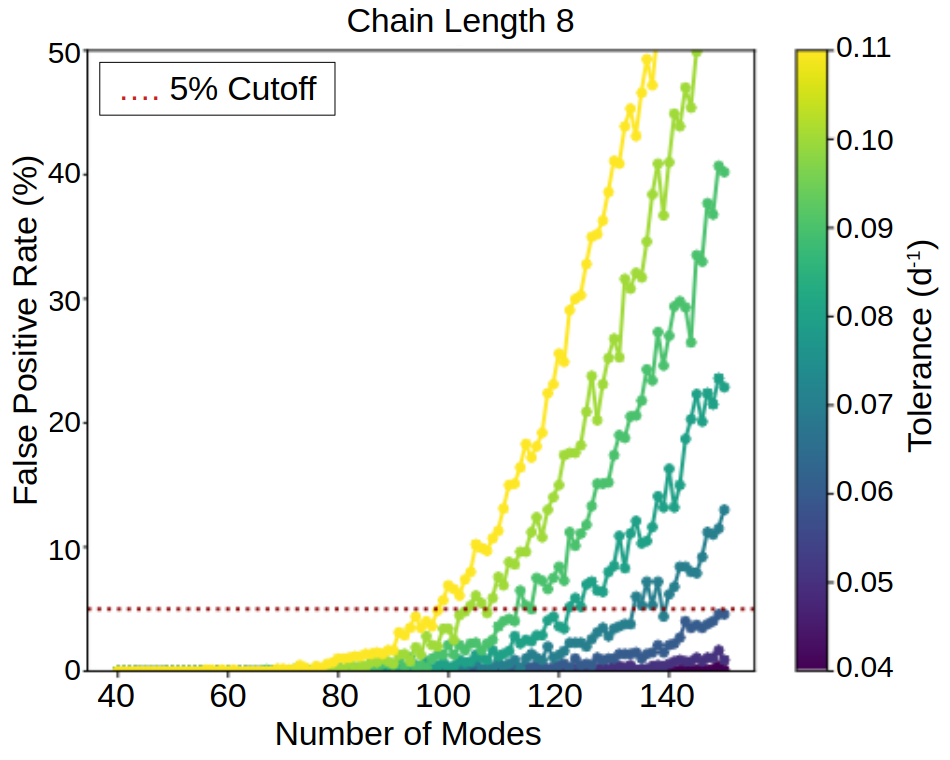} 
    \caption{
    False Positive Rate (FPR) as a function of the number of peaks for regularly spaced chains of lengths 4, 5, 6, and 8. 
    Each curve corresponds to a specific tolerance (in $d^{-1}$), represented by the color scale. 
    The dashed red line indicates a 5\% FPR cutoff which we have used as a threshold beyond which we deem the chains untrustworthy. 
    The FPR increases with both the number of modes and the tolerance. 
    }
    \label{fig.fpr}
\end{figure*}

The FPR is then quantified as the fraction of trials yielding at least one valid chain detection. In Figure~\ref{fig.fpr}, we plot the FPR for fixed chain length as a function of tolerance and the number of detected peaks in the spectrum. As the chain length reduces, the likelihood of finding a random AAP of this length among a sequence of frequencies increases, thereby leading to decreased trust. Chains of length 8 are for instance much harder to detect than those of length 4. We show the FPRs as function of the number of peaks, chain lengths and tolerances in Figure~\ref{fig.fpr}.

Lastly, we note that some caution is necessary in interpreting the FPR. Although we compute FPRs for AAPs using Monte Carlo–generated random sequences, we emphasize that this is only a statistical baseline. In observed spectra, the detection of such sequences is physically meaningful: the peaks represent stellar oscillation modes, not arbitrary numbers. Thus, FPRs derived from random sequences should not be applied mechanically; they provide a useful yardstick, but the astrophysical context gives an additional (unquantifiable) weight to an observed sequence that synthetic calculations cannot replicate.

{
\subsection{Assessment on Synthetic Spectra}
Before applying the algorithm to observed spectra, we examine how accurately it can recover the already-known large spacings of numerically computed spectra. We present here the spectral analysis of a chemically homogeneous $2M_\odot$ star {uniformly rotating at a rate 20\% of its Keplerian breakup (given by $\Omega_K = \sqrt{GM_\star/R_{\star\rm eq}^3}$).}
The basic {2D} 
structure of the star {is computed using the Self Consistent Field (SCF) method} and the effect of rotation is fully taken into account in the mode calculations \citep{Reese2009}.} 
We restricted the scanning of acoustic frequencies to the range $[2.5-12.584]\Omega_K$, consistent with typical p--mode frequencies observed in $\delta$ Scuti stars. The large-frequency separation $\Delta\nu$ was determined to be $7.1036 {\rm d}^{-1}$ based on the frequencies of {2-period island modes $\ell=0-1$, $m=0$ and $n=4-8$.} \\
 
\begin{figure*}
    \centering
    \includegraphics[width=\linewidth]{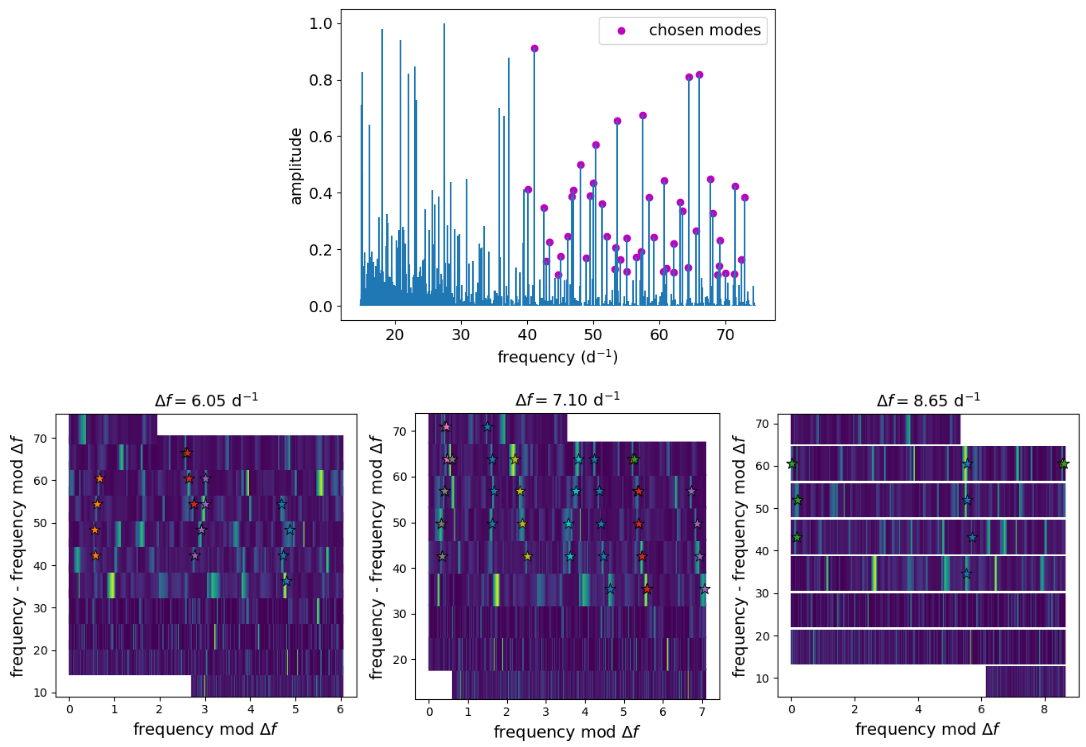}
    \caption{Theoretical pulsation spectrum of a $2M_\odot$ $\delta$ Scuti star (top) with amplitude randomized following a log-uniform distribution. The \'echelle diagrams for three different regular frequency separations (bottom panel) with the stars symbols annotated to highlight the vertical ridges.}
    \label{fig:scf_60_1}
\end{figure*}

{Modes are assigned amplitudes
corresponding to mode visibility at a 60$^\circ$ inclination angle multiplied by a random number sampled from a log-uniform distribution, where logarithms are drawn from the range  0 and 1. This is an intent to account for the apparent randomness of mode-excitation mechanisms.} \\

\ \\
{
We chose the 150 highest amplitude modes (Figure \ref{fig:scf_60_1}: top panel) and applied our algorithm to detect among them sequences consisting of at least 4 equally-spaced frequencies. For different values of candidate $\Delta f$ values, the algorithm was able to find multiple sequences, each comprising nearly equally separated frequencies. The \'echelle diagrams (Figure \ref{fig:scf_60_1}: bottom panel) illustrate regular ridges detected for three distinct $\Delta f$, namely 6.05, 7.10, and 8.65 ${\rm d}^{-1}$. The $\Delta f$ $7.10 ~{\rm d}^{-1}$ detected by our algorithm is close to the large spacing  $\Delta\nu = 7.1036 {\rm d}^{-1}$. While the number of sequences detected for far-off $\Delta f$ values are low, and sometimes overlapping, the sequences found for $\Delta f$ closer to the correct $\Delta\nu$ are distinct (lower extent of overlapping) and their number is maximized. Of the many sequences found for $\Delta f = 7.10 ~{\rm d^{-1}}$, one has frequencies (40.13850396, 47.06793808, 54.11661677, 61.17416694, 68.14355421) ${\rm d^{-1}}$, which are close to the model-identified 2-period island modes $\ell=2, m=2, n=4-8$.
}

\section{Results}

\begin{figure*}[!h]
    \centering
    \includegraphics[width=\linewidth]{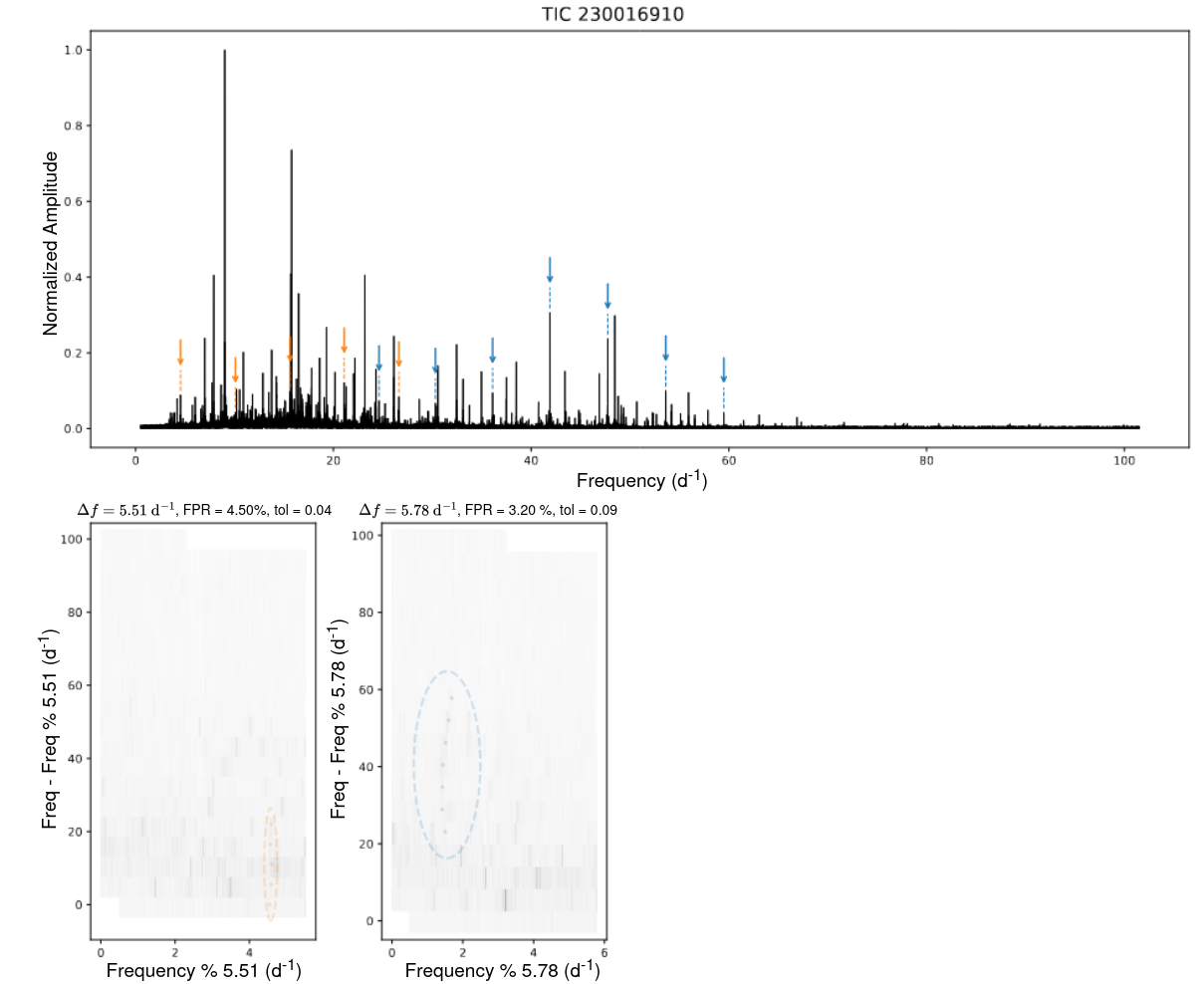}
    \caption{
    Spectra and echelle diagrams displayed together for multiple regular frequency separations ($\Delta f$). 
    Each echelle is labeled with its corresponding tolerance (tol), false positive rate (FPR), and $\Delta\nu$. 
    Arrows on the spectra indicate detected modes which belong to a certain sequence, while the colored contours on the echelle diagrams show the ridges to which these modes belong. 
    Both arrows and contours are color-coded to enhance visual correspondence between the spectra and echelle features.
    More such plots are provided in the {Appendix \ref{A:ech}}.
    }
    \label{fig-spectra1}
\end{figure*}

Automating the algorithm allows us to apply it in a straightforward manner and examine the 4,681 $\delta$ Scuti stellar oscillation spectra. Of these, our algorithm identifies 2,567 spectra (55\%) with least one regular sequence at high statistical significance. {Nonetheless, instances with multiple identified sequences could possibly be a more definitive indication of island-mode patterns.}
{Figure \ref{fig-spectra1} shows an instance where regular frequency spacings have been detected by our automated algorithm. Appendix \ref{A:ech} presents a few more examples of spectra with \ech diagrams containing regular sequences at multiple spacings, marked by faint contours to draw the reader's eye without being distracting. The TIC IDs, detected regular spacings along with stellar parameters for the stars analyzed here are summarized in Table \ref{tab:stellar-params} of Appendix \ref{A:table1}.
}

\begin{figure*}[!h]
    \centering
    \includegraphics[width=0.9\linewidth]{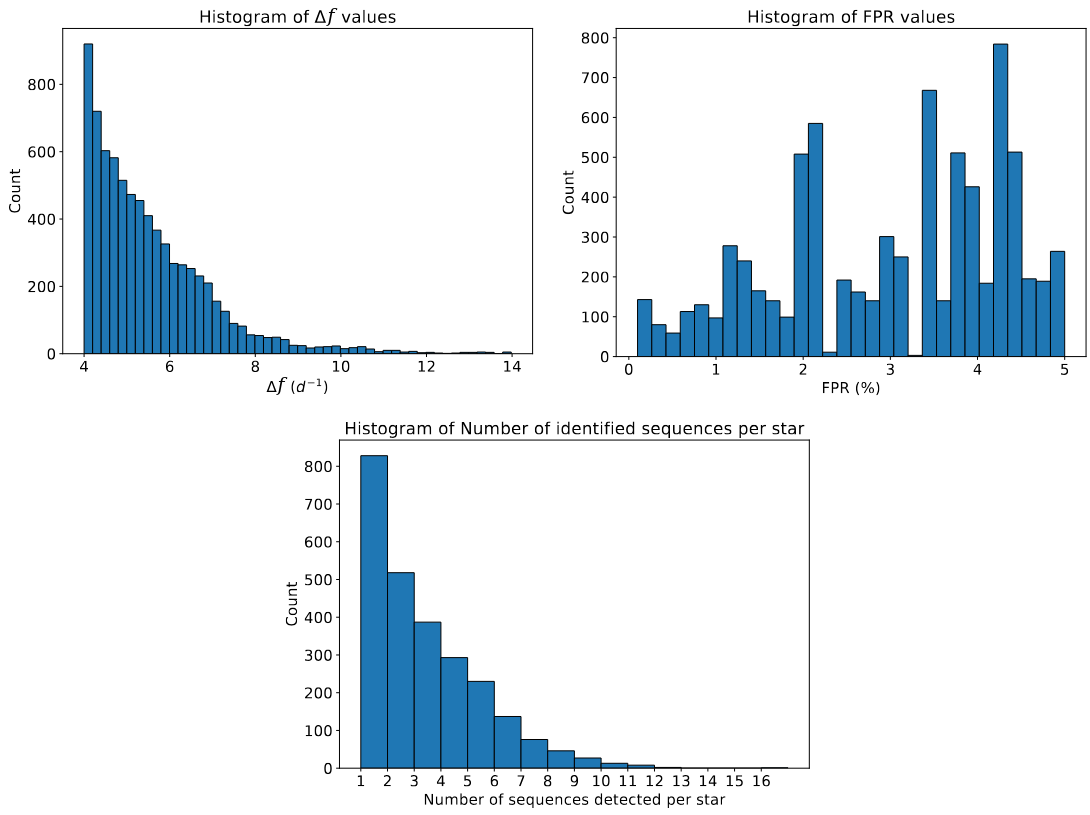}
    \caption{
    Histograms showing identified $\Delta f$ values, false positive rates (FPR), 
    and the number of $\Delta f$ sequences detected per star for the TESS sample of $\delta$ Scuti stars.
    }
    \label{fig.hist}
\end{figure*}

A histogram showing the identified spacings, a distribution of the FPRs and the number of sequences is presented in Figure~\ref{fig.hist}.
This yield rate of $\sim55\%$ for spectra with at least one regular sequence is higher than the $\sim6$\% rate reported in prior studies \citep{Bedding_2020, singh2025seismicconstraintsspinevolution}. The reason for the significantly upwardly revised estimate is the inclusion of low-amplitude peaks in the analysis, for which our algorithm allows. Previously identified sequences showed strong signatures in their \'{e}chelle diagrams whereas the present detections can contain large amplitude contrasts within individual sequences.

{Mode amplitudes are an important factor in influencing the detectability of regular frequency patterns. These depend on the inclination angles of the spin axes with respect to the observer and, more sensitively, to the non-uniform nature of mode excitation. Another factor is the presence of high-frequency modes in the spectrum: regularity in mode spacing is an outcome of asymptotic theory and is only expected for high radial orders ($n\gtrsim4$). 
 
We also enforce the rule that all the modes in the sequence ought to be detectable at the specified SNR; if we were to allow for the occasional missing mode in a sequence, the detection yield would likely be higher. Finally some of the detection} may be attributed to false positives: a fraction of these sequences probably are just random occurrences of periodic frequencies. However, it is difficult to say which are trustworthy or not at this level of analysis. 

\begin{figure*}[!h]
    \centering
    \includegraphics[width=\linewidth]{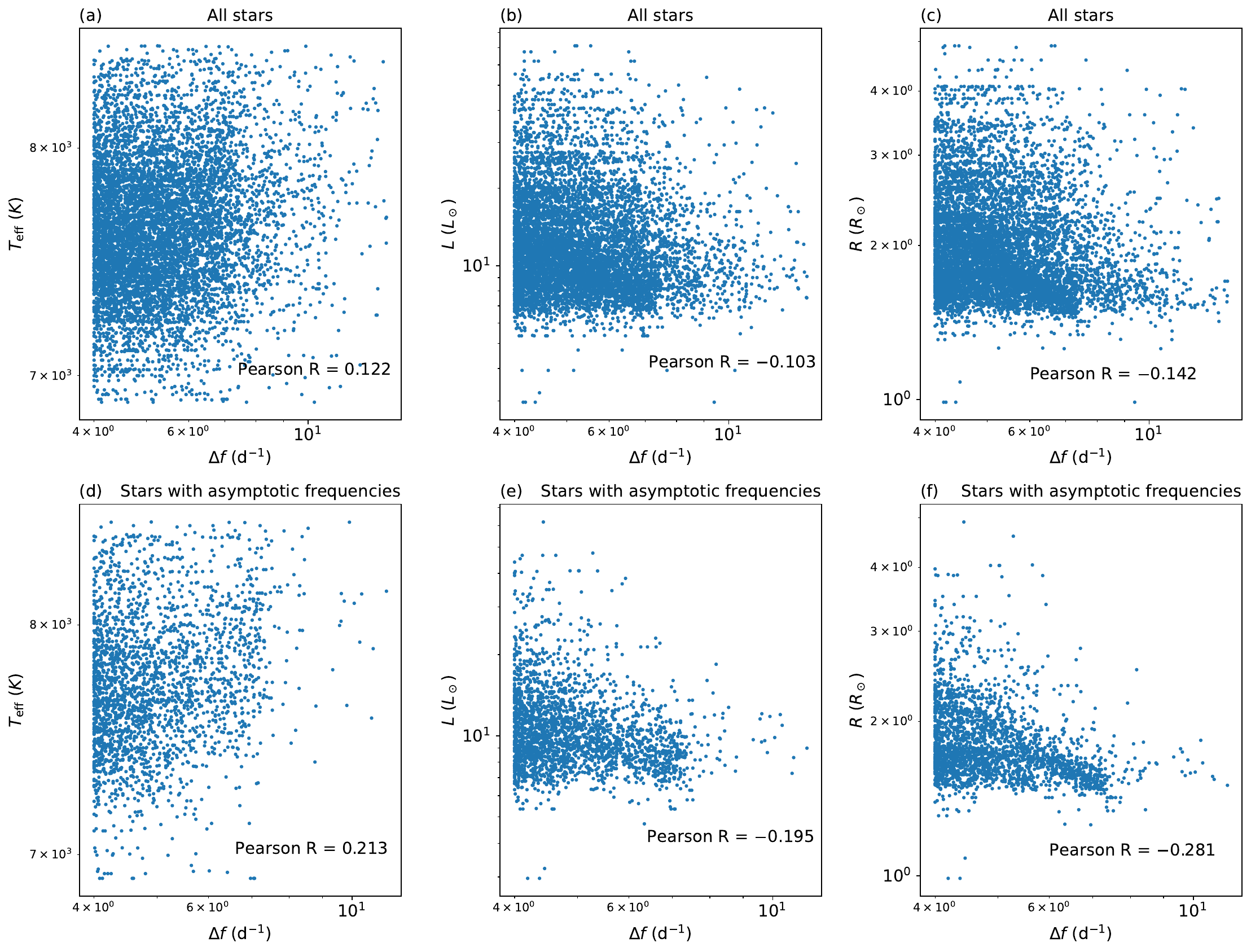}
    \caption{No discernible correlations are seen in the effective-temperature (or luminosity {{, and radii}}) vs. large-spacings plot from TESS observations. Top panels ({{a,b,c}}): -- we take into consideration all the identified regular frequency spacings, i.e., some stars have multiple spacings and we plot all four values here, with the intent to discover possible correlations between some subset of the spacings. The lack of correlation could point to some to some of these sequences being unrealistic or the mode paths being insensitive to the core (which is related to temperature {as H-fusion at the center plays an important role in setting the structure of the star}). Bottom panel: {({d,e,f}}): -- we only plot limited $\Delta f$ for selected stars whose \'echelle diagrams exhibit nearly vertical ridges comprising at least four modes with frequencies above six times the $\Delta f$ asserting their asymptotic nature. The datasets used to create this figure are available in machine-readable format in the online journal.}
    \label{temperature-luminosity}
\end{figure*}

{Since the large separation scales with the square root of the star mean density it decreases with the evolutionary stage of the star. Thus, we expect to see a meaningful correlation between large spacings and temperature / luminosity; however, virtually no correlation is seen in the upper panels of Figure~\ref{temperature-luminosity}. These top panels contain all the detected spacings (sometimes multiple values for each star). To refine this experiment further, we focus on the high-frequency regime by only retaining}
spacings where at least some of the peaks are at frequencies $\lfloor\nu/\Delta f\rfloor \ge 6$ and plot them against the luminosities and effective temperatures of the stars in the lower panels of Figure~\ref{temperature-luminosity}. This leads to a slight improvement in the level of correlation.

\section{Circuit-completion times of 2- and 6-period island modes}
{
Because of rotational deformation, the acoustic rays propagate along different circuits, possessing varied path lengths. Consequently, it is possible that the sound crossing times for these 2- and 6-period island-mode (and possibly others) circuits varies considerably, leading to a range of differing frequency spacings $\Delta$. We computed a 2D baroclinic structure of a 2$M_\odot$ star rotating at 0.2 $\Omega_K$ using the ESTER code and tracked the dimensionless acoustic travel times for two different mode types in units of $\tau_{\rm ref} = \sqrt{R^3/GM_\star}$.}

\begin{figure*}
\gridline{
\fig{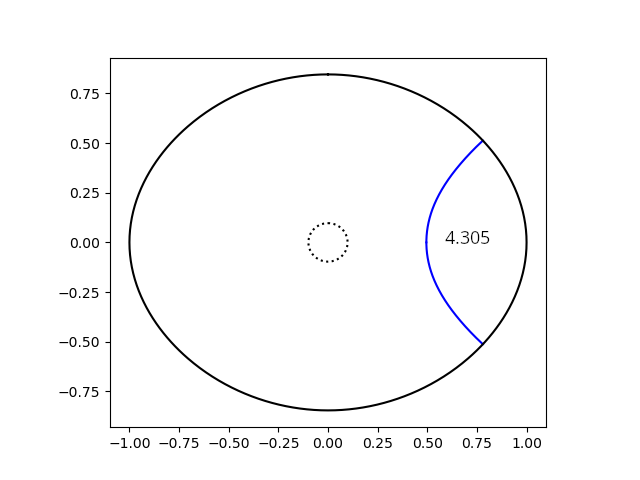}{0.34\textwidth}{(a)}
\fig{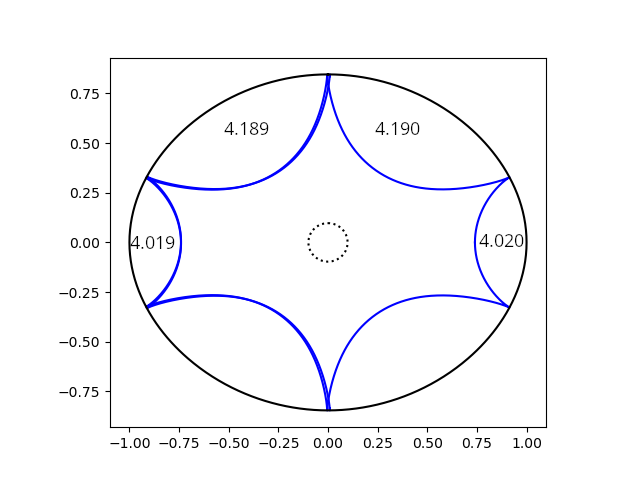}{0.34\textwidth}{(b)}
\fig{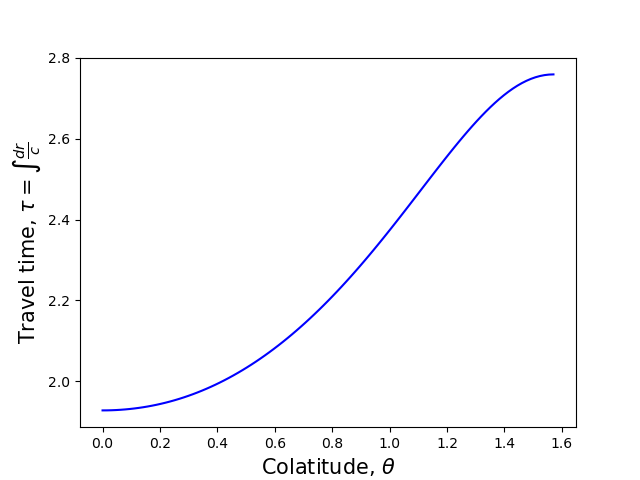}{0.34\textwidth}{(c)}
}
\caption{
{Examining the sound crossing times in dimensionless units ($\tau_{\rm ref}$) for different island modes propagating and bouncing between different colatitudes. The ESTER code was used to produce 2D baroclinic models of a 2$M_\odot$ star rotating at 0.2 $\Omega_K$. (a) Ray path for a 2-period island mode with sound travel time of 4.305 $\tau_{\rm ref}$. (b) Similar diagram for a 6-period island mode with slightly varying bouncing time, i.e., 4.02 and 4.19 $\tau_{\rm ref}$. (c) Acoustic travel time as function of colatitude $\theta$: 0 represents the pole and $\pi/2$ the equator.}
}
\label{fig:soundtime}
\end{figure*}

{The acoustic travel time of the 2-period island mode (Figure \ref{fig:soundtime}-a) is around 13105 seconds (4.30 $\tau_{\rm ref}$). Similarly the sound crossing time for the 6-period island modes (Figure \ref{fig:soundtime}-b) were either 12234 seconds (4.02 $\tau_{\rm ref}$) or 12746 seconds (4.19 $\tau_{\rm ref}$). Roughly, the circuit-crossing times of these island modes vary by less than 10\%. Further, the sound crossing time appears to increase with colatitude (Figure \ref{fig:soundtime}-c); however the difference between the polar ($\theta=0$) and equatorial ($\theta= \pi/2$) mode is not so large that it would result in variations beyond 10\% in frequency spacings $\Delta f$. The evidence from this star suggests that the circuit dependence of $\Delta f$ as a hypothesis may not serve as an explanation for the multiple $\Delta f$ values we detected.}

\section{Implications for Mode Organization in $\delta$ Scuti Stars}
The principal results of this work are twofold. First, we present an automated statistical framework for identifying regular mode sequences in large stellar oscillation datasets while quantifying the significance of each candidate detection. Second, applying this framework to 4,681 TESS $\delta$ Scuti stars reveals that regular mode sequences are common and that many stars exhibit multiple independent regular frequency spacings within a single oscillation spectrum. Together, these findings demonstrate that automated survey-scale analyses can uncover regularities that are difficult to identify through manual inspection of individual spectra.

The algorithm that we introduce here is not merely a computational convenience; it enables a qualitatively different view of $\delta$ Scuti oscillation spectra by allowing statistically controlled, and computationally cheap searches across thousands of stars. Worth highlighting is that, while we provide a starting point for characterizing the false positive rate, the technique itself requires further refinement, {such as implementation of variable tolerance and local SNR}. Variable tolerance is especially useful in accounting for ridge curvature, which typically affects lower-frequency regions of the oscillation spectrum.

Previous studies concentrated primarily on the highest-amplitude oscillation modes and relied heavily on manual inspection of \'{e}chelle diagrams. The present algorithm searches systematically for weaker sequences while assigning a statistical confidence to each candidate, making it practical to analyze thousands of spectra in a homogeneous fashion. Although regular sequences are not recovered in every spectrum, non-detections may arise from limited signal-to-noise, incomplete mode visibility, intrinsically more complex pulsation spectra, or limitations of the present detection framework. Longer observational baselines from additional TESS sectors and future missions such as PLATO should improve sensitivity to weaker oscillation modes and enable more complete seismic analyses.

One possible explanation for the multiple regular mode sequences identified here is that they arise from different classes of acoustic modes predicted by ray-dynamical theory. Existing calculations indicate that island and chaotic mode families should produce regular frequency sequences with characteristic spacings close to $\Delta f$ or rational fractions such as $\Delta f/2, \Delta f/3$ \citep{evano19}. To examine whether differences in acoustic travel times between these mode families could account for the diversity of spacings observed in individual stars, we computed representative ray paths and their associated sound-crossing times in a two-dimensional ESTER model. The resulting travel times differ by less than approximately 10\%, substantially smaller than the differences between many of the spacings recovered in the observations. While this comparison is necessarily limited to a single stellar model, it suggests that variations in the acoustic path lengths of currently identified island-mode families alone are unlikely to explain the full diversity of regular spacings. Other possibilities include additional mode families that have not yet been explored theoretically, the effects of stellar evolution and rotation across a broader region of parameter space, or physical processes that are not fully captured by current calculations.

The primary observational result is the frequent occurrence of multiple independent regular frequency spacings within the same oscillation spectrum. Existing theoretical calculations predict that high-frequency acoustic modes should organize into regular sequences associated with different ray-dynamical mode families, with characteristic spacings close to $\Delta$, $\Delta/2$, and under limited circumstances $\Delta/3$. These predictions account naturally for the existence of regular frequency sequences, but do not readily explain the diversity of spacings recovered in many of the observed spectra.

The present results provide new observational constraints on theories of oscillations in rapidly rotating $\delta$ Scuti stars. Whether the observed diversity of regular spacings reflects additional mode families, incomplete exploration of stellar parameter space, or other physical effects remain open questions. Future theoretical calculations spanning a wider range of masses, evolutionary states, and rotation rates, together with longer-duration photometric observations, will be essential for determining the physical origin of these multiple regular mode sequences.

Alternately, if we were to relax this constraint and allow at least one mode within the sequence to possess a local SNR of 1 (i.e., undetected), the detection rate would likely rise, leading to many more ridge identifications. {{We describe the methodology and report some detections in the Appendix.}}
Longer observational baselines, such as with successive TESS sectors or the forthcoming PLATO mission, may improve the SNR levels of these pulsators.

 One departure from solar-like oscillators is the detection of multiple regular spacings for {a significant}
 fraction of $\delta$ Scuti stars. {As detailed in the introduction, the asymptotic theory of stellar oscillations and the linear computations of acoustic modes in rapidly rotating stars predict that high frequency modes should exhibit regular spacings close to the large separation.
 By close, we mean that the spacings can differ by a few percent at most \citep{evano19}. This is at odds with the differences between the spacings we detect in stars with  multiple regular spacings. 
 One possibility would be that such multiple spacings exist in theoretical spectra but have not been detected. In particular, the spectra of high-frequency island and chaotic modes can be viewed as being organized in series of regular frequency sequences with spacings close to large separations $\nu_{n,i} = \Delta_i(n + \epsilon_i)$ \citep{evano19}. Except for some island modes, these series are expected to be weakly correlated, i.e., $\epsilon_i$ may be modeled as a random number between $0$ and $1$. For a sufficiently large number of modes in a sample, such an organization will produce multiple spacings in addition to $\Delta$. The non-uniform excitation mechanism plays an important role in determining the excitation and detection of sequences in observed spectra with at least $6$ equally spaced frequencies.
 One possibility is that we are missing some important physical ingredients. For instance, theoretical acoustic spectra have been analyzed in detail only for a few stellar structure models. In particular, regular spacings in the spectra of evolved stars in the $\delta$ Scuti stellar mass range have not been investigated yet.} 

\begin{acknowledgements}
The authors thank Dr. Fran\c{c}ois Ligni\`{e}res for valuable discussions and his theoretical insights. SH acknowledges support from the Department of Atomic Energy, Government of India, under Project Identification No. RTI 4002. This research was supported in part by generous donations from the Murty Trust and Premji Invest, both aimed at enabling advances in astrophysics through the use of machine learning. Murty Trust, an initiative of the Murty Foundation, is a not-for-profit organisation dedicated to the preservation and celebration of culture, science, and knowledge systems born out of India. The Murty Trust is headed by Mrs. Sudha Murty and Mr. Rohan Murty.
\end{acknowledgements}
\appendix

\section{\'Echelle diagrams} \label{A:ech}

A complete set of \'echelle diagrams for some of the spectra in which we detected sequences is displayed in this section. The remaining spectra along with the \'echelle diagrams may be found at \dataset[DOI: 10.5281/zenodo.22064659]{https://doi.org/10.5281/zenodo.22064659}.

\begin{figure*}[!h]
    \centering
    \includegraphics[width=\linewidth]{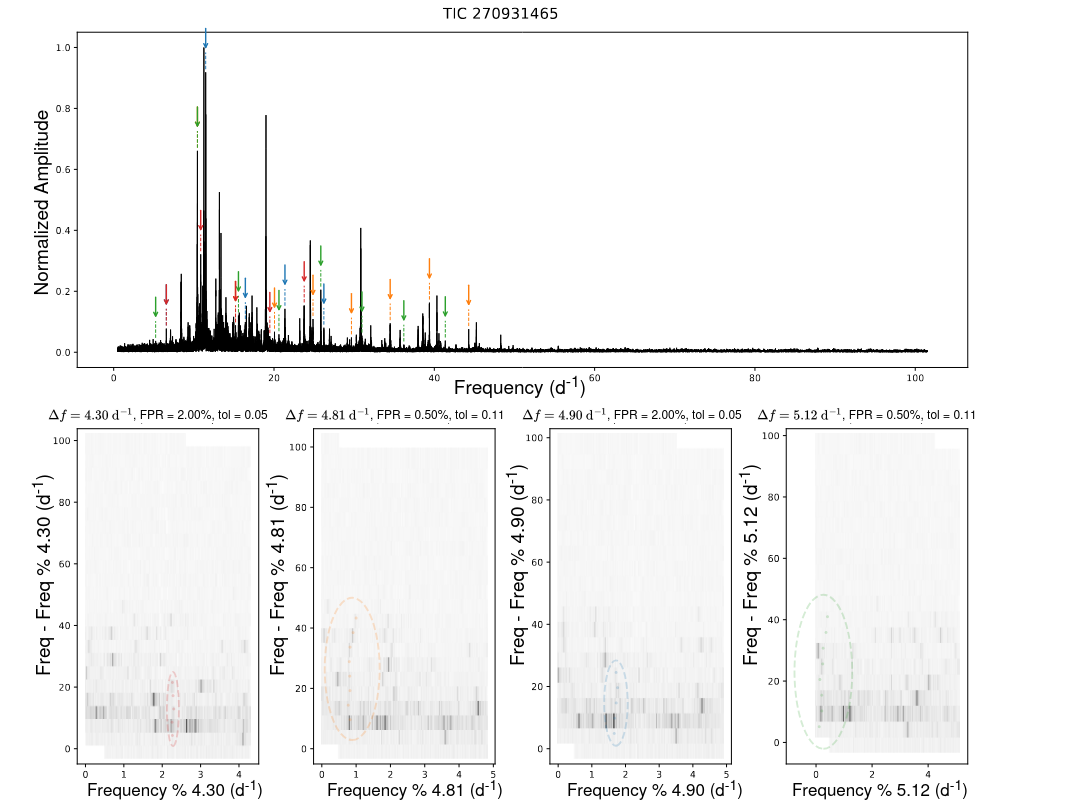} 
    \caption{
    Spectra and \'echelle diagrams displayed together for multiple regular frequency separations ($\Delta f$). 
    Each \'echelle is labeled with its corresponding tolerance (tol), false positive rate (FPR), and $\Delta f$. 
    Arrows on the spectra indicate detected modes which belong to a certain sequence, while the colored contours on the \'echelle diagrams show the ridges to which these modes belong. 
    Both arrows and contours are color-coded to enhance visual correspondence between the spectra and \'echelle features.
    }
    \label{fig-spectra2}
\end{figure*}

\begin{figure*}[!h]
    \centering
    \includegraphics[width=\linewidth]{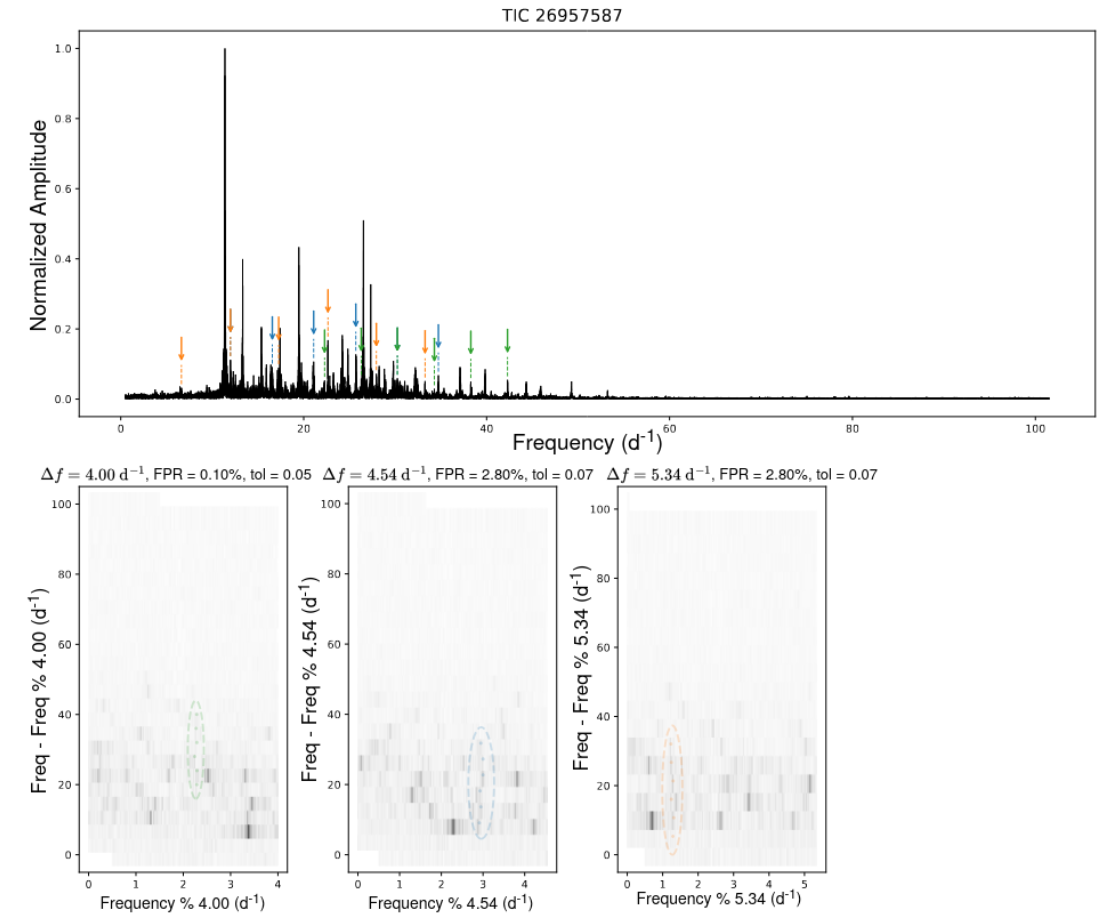}   
    \caption{
    Spectra and echelle diagrams displayed together for multiple regular frequency separations ($\Delta f$). 
    Each echelle is labeled with its corresponding tolerance (tol), false positive rate (FPR), and $\Delta f$. 
    Arrows on the spectra indicate detected modes which belong to a certain sequence, while the colored contours on the echelle diagrams show the ridges to which these modes belong. 
    Both arrows and contours are color-coded to enhance visual correspondence between the spectra and echelle features.
    }
    \label{fig-spectra3}
\end{figure*}

\begin{figure*}[!h]
    \centering
    \includegraphics[width=\linewidth]{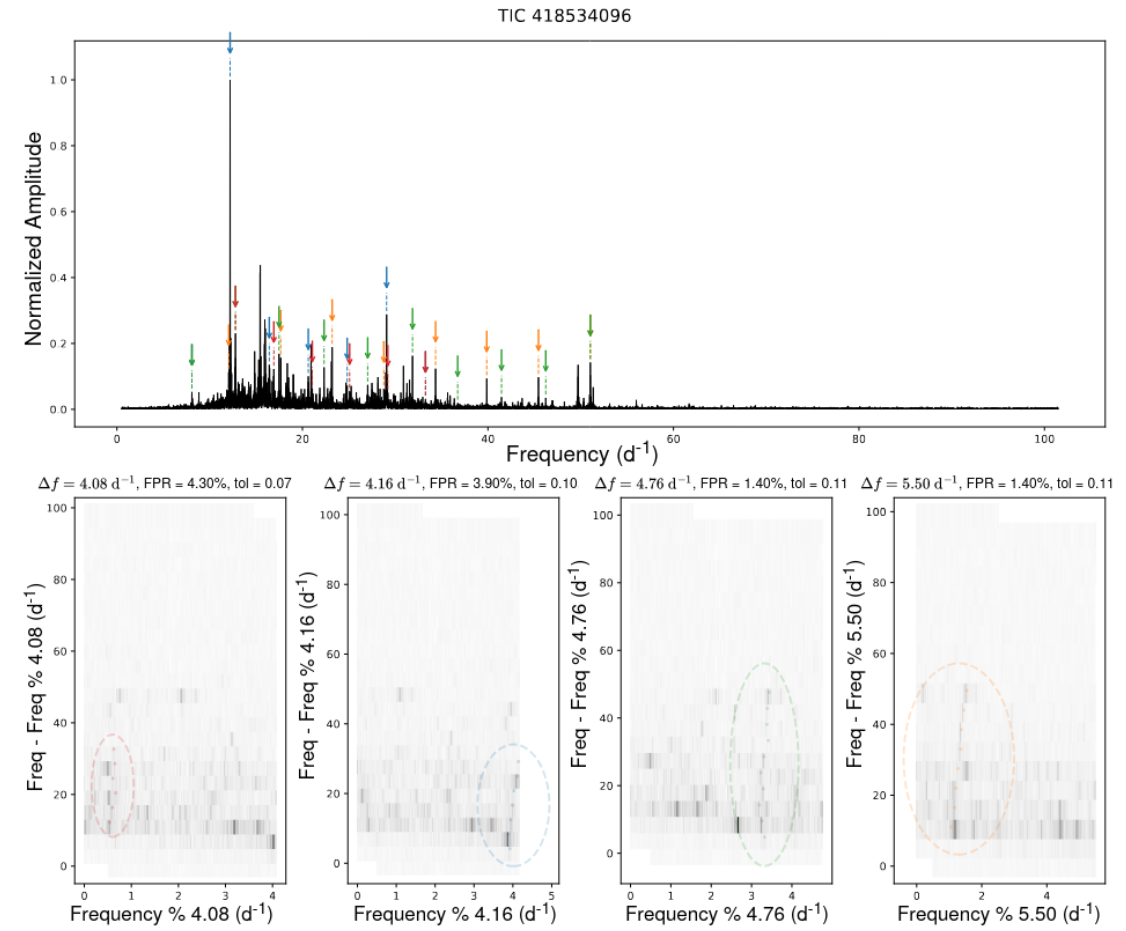}  
    \caption{
    Spectra and \'echelle diagrams displayed together for multiple regular frequency separations ($\Delta f$). 
    Each \'echelle is labeled with its corresponding tolerance (tol), false positive rate (FPR), and $\Delta f$. 
    Arrows on the spectra indicate detected modes which belong to a certain sequence, while the colored contours on the \'echelle diagrams show the ridges to which these modes belong. 
    Both arrows and contours are color-coded to enhance visual correspondence between the spectra and \'echelle features.
    }
    \label{fig-spectra4}
\end{figure*}

\begin{figure*}[!h]
    \centering    
    \includegraphics[width=\linewidth]{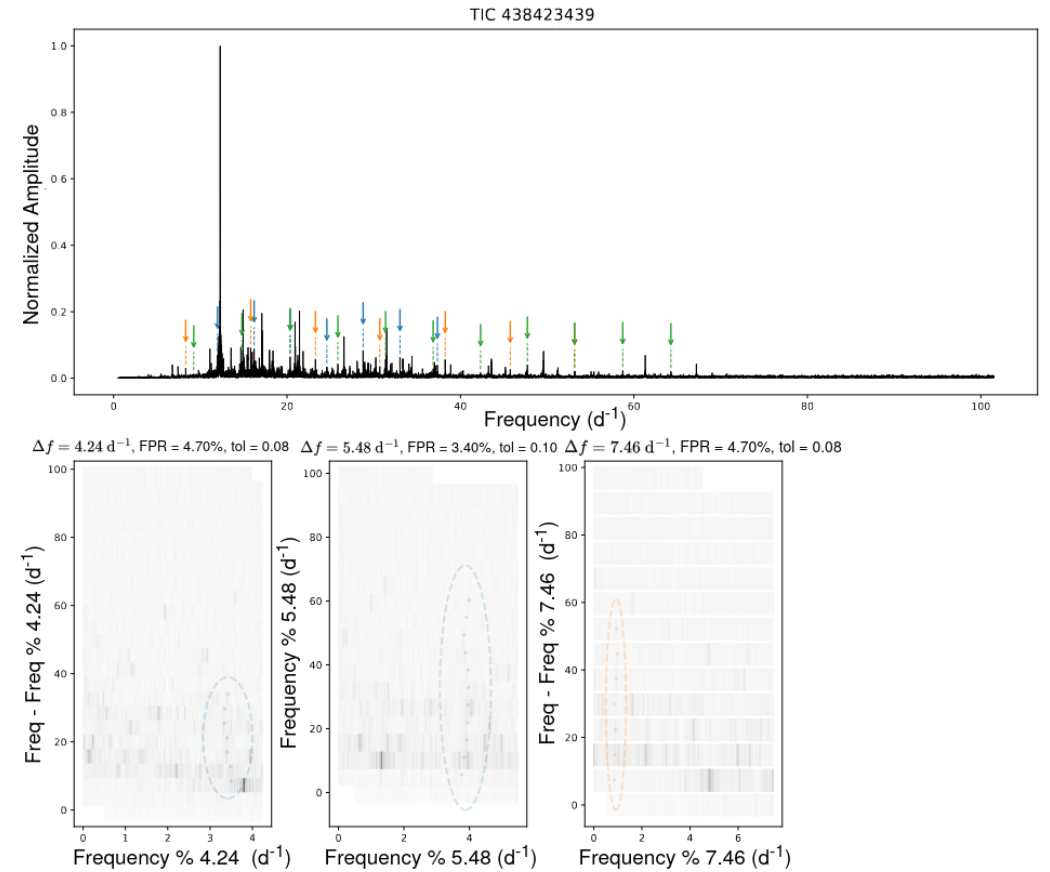}    
    \caption{
    Spectra and echelle diagrams displayed together for multiple regular frequency separations ($\Delta f$). 
    Each echelle is labeled with its corresponding tolerance (tol), false positive rate (FPR), and $\Delta f$. 
    Arrows on the spectra indicate detected modes which belong to a certain sequence, while the colored contours on the echelle diagrams show the ridges to which these modes belong. 
    Both arrows and contours are color-coded to enhance visual correspondence between the spectra and echelle features.
    }
    \label{fig-spectra5}
\end{figure*}

\begin{figure*}[!h]
    \centering
    \includegraphics[width=\linewidth]{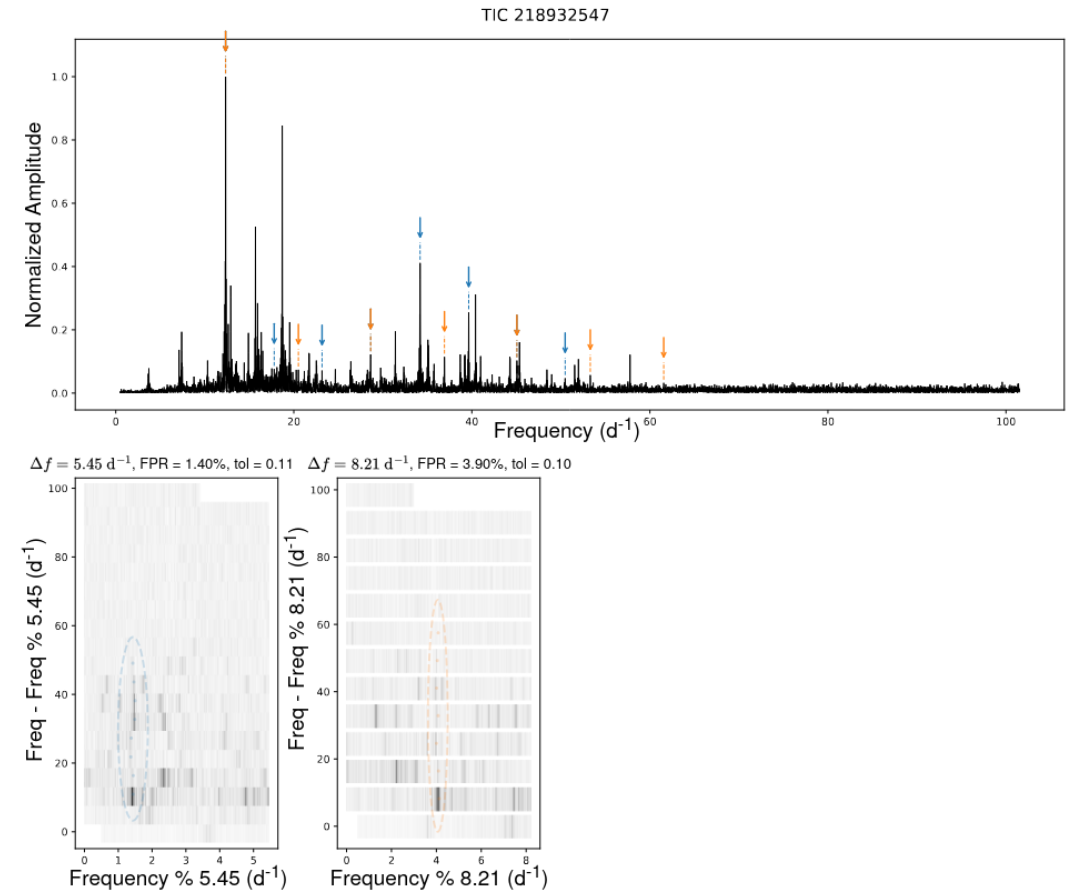}  
    \caption{
    Spectra and \'echelle diagrams displayed together for multiple regular frequency separations ($\Delta f$). 
    Each \'echelle is labeled with its corresponding tolerance (tol), false positive rate (FPR), and $\Delta f$. 
    Arrows on the spectra indicate detected modes which belong to a certain sequence, while the colored contours on the \'echelle diagrams show the ridges to which these modes belong. 
    Both arrows and contours are color-coded to enhance visual correspondence between the spectra and \'echelle features.
    }
    \label{fig-spectra6}
\end{figure*}

\begin{figure*}[!h]
    \centering
    \includegraphics[width=\linewidth]{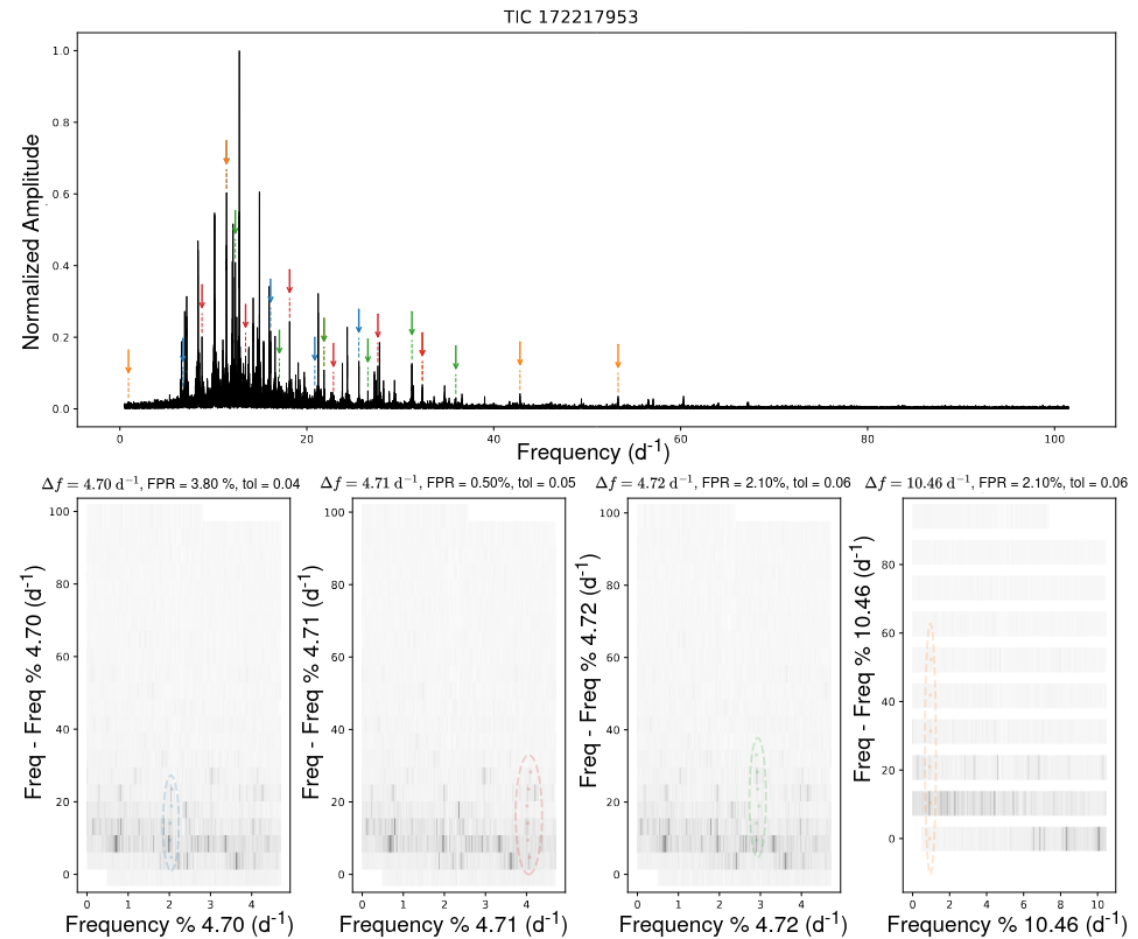}  
    \caption{
    Spectra and \'echelle diagrams displayed together for multiple regular frequency separations ($\Delta f$). Each \'echelle is labeled with its corresponding tolerance (tol), false positive rate (FPR), and $\Delta f$. 
    Arrows on the spectra indicate detected modes which belong to a certain sequence, while the colored contours on the \'echelle diagrams show the ridges to which these modes belong. 
    Both arrows and contours are color-coded to enhance visual correspondence between the spectra and \'echelle features.
    }
    \label{fig-spectra7}
\end{figure*}

\clearpage

\section{Detected frequency spacings and stellar parameters} \label{A:table1}

Table \ref{tab:stellar-params} presents the TIC IDs, detected multiple regular frequency spacings and the stellar parameters derived for some of the stars analyzed in this work. The $\Delta f$ values and corresponding regular frequency sequences are compiled in a machine-readable table that is available in the online journal (see Figure \ref{temperature-luminosity}).
Physical parameters such as luminosity, effective temperature, and radii of the analyzed targets are also provided in a machine-readable table that is available in the online journal (see Figure \ref{temperature-luminosity}).

\begin{table}[H]
\centering
\caption{Stellar parameters including selected frequency spacings ($\Delta f$), FPR, luminosity, effective temperature, mass, and radius values for some stars.}
\label{tab:stellar-params}
\resizebox{\textwidth}{!}{
\begin{tabular}{ccccccccc}
\toprule
TIC &
$\Delta f_1$ (d$^{-1}$) &
FPR$_{1}$ (\%) &
$\Delta f_{2}$ (d$^{-1}$) &
FPR$_{2}$ (\%) &
Luminosity (L$_\odot$) &
$T_{\rm eff}$ (K) &
Mass ($M_\odot$) &
Radius ($R_\odot$) \\
\midrule
7808834   & 6.46   & 3.4    & 4.88   & 2.7    & 29.74 & 7144    & 1.59  & 3.56 \\
364399376 & 4.45   & 0.6    & 4.22   & 1.9    & 32.49 & 7471    & 1.72  & 3.40 \\
270265624 & 6.08   & 2.4    & 7.15   & 2.4    & 16.13  & 6942    & 1.52 & 2.78 \\
191466237 & 5.05   & 4.8    & 7.79   & 3.7    & 23.83 & 7057 & 1.56  & 3.26 \\
139825582 & 4.06   & 2.0    & 5.31   & 4.5    & 30.23  & 7244    & 1.63  & 3.50   \\
102204751 & 5.84   & 3.4    & 5.37   & 3.4    & 25.87   & 7543    & 1.74 & 2.98 \\
300808209 & 5.27   & 3.7    & 4.72   & 3.7    & 18.94 & 7226    & 1.62  & 2.78 \\
220478230 & 4.43   & 1.9    & 4.99   & 3.6    & 43.93 & 7409    & 1.69  & 4.02 \\
382341346 & 6.38   & 4.5    & 4.18   & 4.5    & 6.75 & 7428    & 1.70   & 1.57 \\
13708886  & 4.00   & 0.3    & 4.08   & 1.2    & 18.41 & 7179    & 1.60 & 2.77 \\
230016910 & 5.78   & 3.2    & 5.51   & 4.5    & 12.44 & 7912    & 1.90 & 1.88 \\
47205656  & 9.10   & 4.5    & 4.81   & 1.4    & 53.66 & 7452    & 1.71  & 4.39 \\
410447142 & 5.43   & 4.3    & 4.02   & 1.4    & 21.54 & 7437 & 1.70   & 2.79 \\
264253320 & 4.71   & 4.5    & 4.00   & 2.5    & 15.60   & 7043    & 1.55 & 2.65  \\
68149913  & 7.61   & 1.4    & 6.09   & 1.4    & 9.90   & 7790    & 1.85  & 1.73 \\
415333069 & 4.49   & 2.0    & 5.56   & 4.5    & 22.82 & 7303 & 1.65  & 2.98 \\
90322352  & 9.22   & 2.2    & 9.38   & 2.9    & 17.43   & 7397    & 1.69  & 2.54 \\
46882571  & 4.16   & 3.8    & 5.81   & 3.8    & 17.07   & 7176    & 1.60 & 2.67 \\
\bottomrule
\end{tabular}
}
\end{table}

\section{Experiment with Signal-to-noise Ratio}

\begin{figure}
    \centering
    \includegraphics[width=\linewidth]{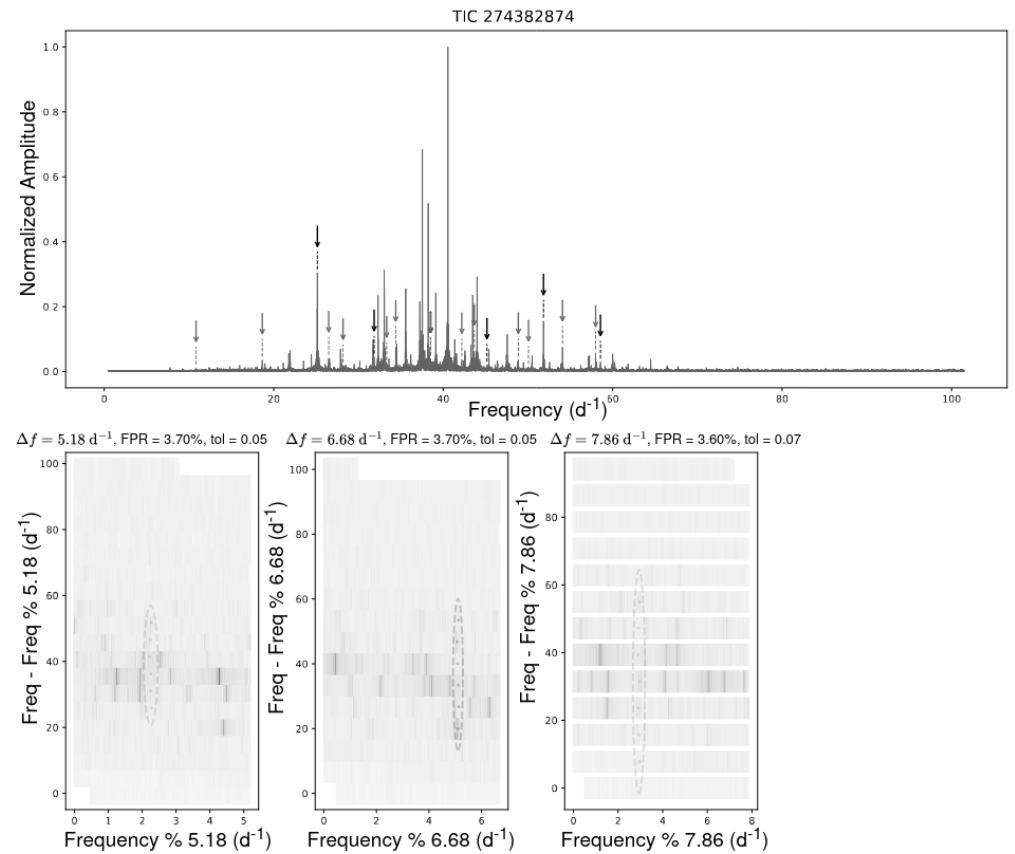}
    \caption{(top): Spectrum of TIC 274382874 and marked frequencies that form coherent ridges in the \'echelle diagrams with different $\Delta f$. (bottom): Shown are the corresponding \'echelle diagrams, overlaid with the regular frequency patterns annotated. The tolerance within which the frequencies are equally spaced, and statistical false positive rates (FPR) indicating the randomness probability for such occurrence are mentioned in the title.}
    \label{fig:relaxed_1}
\end{figure}

\begin{figure}
    \centering
    \includegraphics[width=\linewidth]{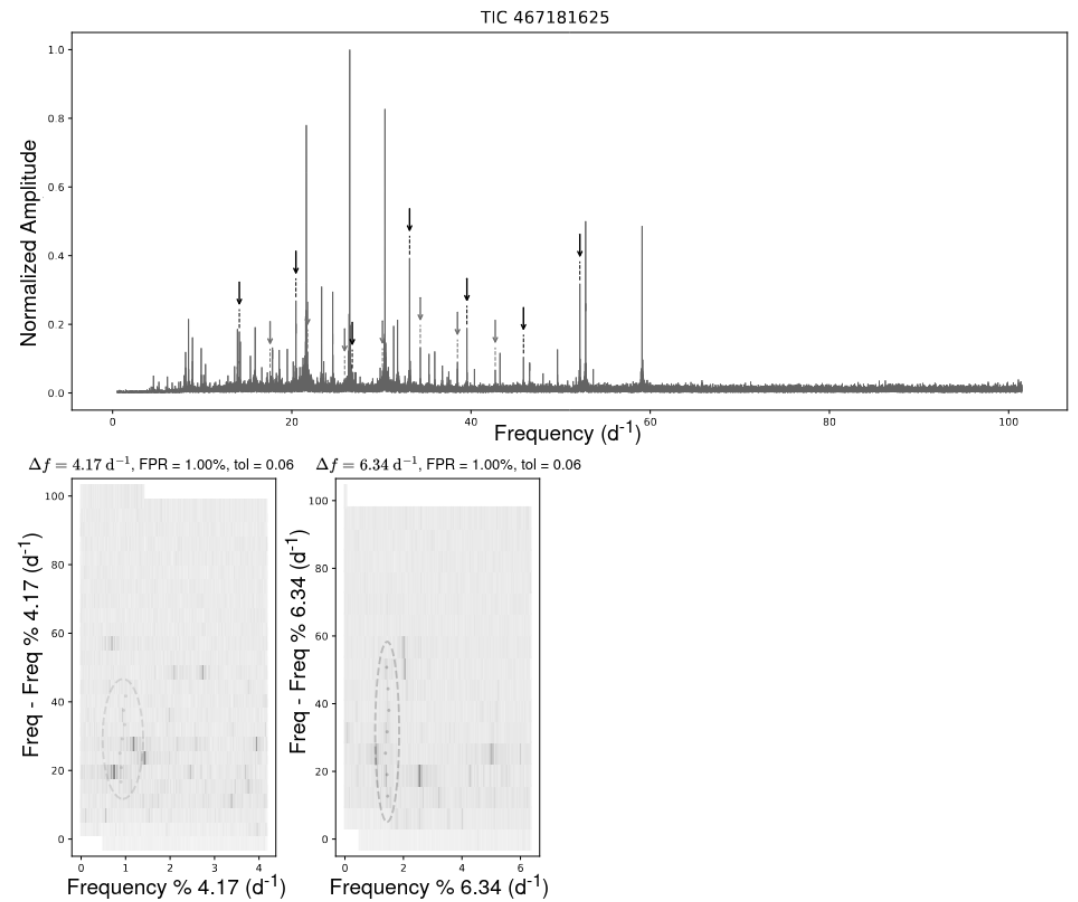}
    \caption{(top): Spectrum of TIC 467181625 and marked frequencies that form coherent ridges in the \'echelle diagrams with different $\Delta f$. (bottom): Shown are the corresponding \'echelle diagrams, overlaid with the regular frequency patterns annotated. The tolerance within which the frequencies are equally spaced, and statistical false positive rates (FPR) indicating the randomness probability for such occurrence are mentioned in the title.}
    \label{fig:relaxed_3}
\end{figure}

{In this section, we explore as a proof of concept, whether regular mode patterns can emerge among the low-amplitude modes in the spectra of two stars for which we were unable to detect regular frequency resonances, perhaps due to modes missing in the middle of an otherwise continuous sequence. The local SNR threshold of 1.2 for recognizing a mode was relaxed to 1.0, allowing a number of extra modes to be included in the graph theory analysis to search for uniform frequency spacing across the spectra. Following the analysis outlined in the main text, we were able to retrieve uniformly-spaced frequency series within smaller tolerance levels in the two stars (see figure \ref{fig:relaxed_1} and \ref{fig:relaxed_3}).}

\end{document}